\documentclass[english,aps,prstper,reprint,showpacs,titlepage,longbibliography] {revtex4-2}   

\usepackage[T1]{fontenc}	
\usepackage[latin9]{inputenc}	
\usepackage{geometry}		

\usepackage{graphicx}
\usepackage[above,below]{placeins}	
\usepackage{times}

\usepackage[abs]{overpic}
\usepackage{pict2e}

\usepackage{hyperref}
\usepackage{breakurl}
\hypersetup{colorlinks=true,urlcolor=blue,citecolor=blue,linkcolor=blue}   
\usepackage{enumitem}          
\setlist{nosep}                 

\usepackage{pdfpages}
\makeatletter
\AtBeginDocument{\let\LS@rot\@undefined}
\makeatother
\begin{document}

\begin{titlepage}

  \title{Towards a Generalized Assessment of Computational Thinking for Introductory Physics Students}

  \author{Justin Gambrell}
  \affiliation{Department of Physics, Drexel University, Disque Hall, 32 S 32nd St., Philadelphia, PA,
    19104} 
  \author{Eric Brewe}
  \affiliation{Department of Physics, Drexel University, Disque Hall, 32 S 32nd St., Philadelphia, PA,
    19104} 


  \begin{abstract}
    Computational thinking in physics has many different forms, definitions, and implementations depending on the level of physics, or the institution it is presented in. In order to better integrate computational thinking in introductory physics, we need to understand what physicists find important about computational thinking in introductory physics.  We present a qualitative analysis of twenty-six interviews asking academic (\textit{N}=18) and industrial (\textit{N}=8) physicists about the teaching and learning of computational thinking in introductory physics courses. These interviews are part of a longer-term project towards developing an assessment protocol for computational thinking in introductory physics. We find that academic and industrial physicists value students' ability to read code and that Python (or VPython) and spreadsheets were the preferred computational language or environment used.  Additionally, the interviewees mentioned that identifying the core physics concepts within a program, explaining code to others, and good program hygiene (i.e., commenting and using meaningful variable names) are important skills for introductory students to acquire. We also find that while a handful of interviewees note that the experience and skills gained from computation are quite useful for student's future careers, they also describe multiple limiting factors of teaching computation in introductory physics, such as curricular overhaul, not having "space" for computation, and student rejection. The interviews show that while adding computational thinking to physics students repertoire is important, the importance really comes from using computational thinking to learn and understand physics better. This informs us that the assessment we develop should only include the basics of computational thinking needed to assess introductory physics knowledge.    \clearpage
  \end{abstract}

  \maketitle
\end{titlepage}

\section{Introduction}

Computation and computational thinking (CT) is currently a growing area of interest due to its applicability across many disciplines. Within physics in particular, there is a synergy between learning discipline-specific content and developing CT skills: physics bolsters CT skills and practices by providing a context rooted in mathematics while CT practice bolsters physics learning by streamlining the process and expanding the physics that is accessible \cite{Guzdial,Basu,Hutchins-1,Hutchins-2, Grover}. Because of this synergy, incorporating CT into the physics degree is a logical outcome. However, curricula in introductory physics often limit their content to problems that can be solved analytically. In order to better serve physics education, therefore, we need to examine how we can update and modernize these curricula to meet the challenges of incorporating CT and maintaining the relevance of physics in the preparation of all students \cite{Undergraduate}.

Multiple researchers and curriculum designers have attempted such reforms by implementing and evaluating CT in physics and other science, technology, engineering, and mathematics (STEM) curricula. Projects and Practices in Physics \((P^3)\), for example, is a curriculum at Michigan State University (MSU) in which students solve highly contextualized problems computationally within small groups. Researchers have examined group interactions in this context~\cite{Obsniuk}, finding that the ways the group interacts with finding bugs in a program are either more strategic by checking line by line, or less strategic by playing with the program until the bug was found. The Partnership for Integration of Computation into Undergraduate Physics (PICUP) is a separate organization that focuses on providing resources for instructors wishing to add computation into their undergraduate physics curricula \cite{PICUP}. Additionally, the American Association of Physics Teachers Undergraduate Curriculum Task Force has released a set of recommendations for computational physics in the undergraduate physics curriculum. Among others, these recommendations include that physics departments should be integrating computation into the curriculum. However, the authors also identified the lack of research on assessment of CT skills as a problem \cite{Undergraduate}.

Thus, the integration of computation into the introductory physics curriculum requires development of assessment protocols for CT that can be used across a variety of classes, but this tool does not yet exist. Our research is working toward the development of such an assessment. Developing assessments is an iterative and cyclical process in which one identifies the criteria to be assessed, gives and evaluates the assessment, and reflects upon the validity of the assessment \cite{Holloman,Wyatt-Smith}.  In this paper, we address the first of these three stages by identifying the learning objectives of a computational introductory physics course. To do so, we conducted interviews with 26 academic and industrial physicists to identify broad learning goals in a computationally integrated introductory physics class, which will guide our assessment development in the future. We specifically aim to answer the following two research questions: \begin{itemize}
    \item \textbf{(RQ1) What are the learning goals for a computationally integrated introductory physics class?}\\ \item \textbf{(RQ2) What do these learning goals suggest for an assessment of computation and computational thinking?}
\end{itemize}


\subsection{What is Computational Thinking?}

 CT is a complicated concept regardless of your experience level with it. How might you describe what thinking is? When does thinking occur or rather when does it not occur?  Thinking (and CT) is esoteric. A definition that varies based on interpretation can be confusing when it comes to scientific research. One such way to address this issue in definition perspective is to provide and explain our own definition which has parts from other definitions provided by Wing and Weintrop:  

\emph{In general, CT is a specialized form of thinking involving problem decomposition, pattern identification/recognition, abstraction of complicated systems along with proper justification, and algorithms brought to bear in order to solve problems that necessitate numeric modeling and problem solving (often involving the use of computers to aid the solving of these problems)\cite{Weintrop,Wing}. Computational literacy is the state of having competent knowledge and use of computational programming environments.  Computational literacy is not needed to perform computational thinking in physics; yet, to manipulate physics programs, some computational literacy is required. Computational thinking in physics is thinking in such a way that computations describing physical phenomena are streamlined, efficient, and well documented, and this is typically executed via programming environment.}

\subsection{CT in introductory physics}

CT is an emerging fundamental skill. It has been a fundamental skill for ages, but only in the past 70 years circa 1950's has it become a specific focus in education \cite{Kong}.  In particular, Jeanette Wing in 2006 popularized the idea of CT and has in part fueled the integration of CT in many disciplines \cite{Wing,Catlin}. We need to be equipping and nurturing every student with the mindset of CT so that it will not only help them obtain jobs in their future, but help develop and promote more ideas that would have been left behind otherwise.  

In PER, computational thinking research is sparse compared to other areas. A paper in 2020 by Odden demonstrates this. In his paper, he thematically codes 18 years (2001-2018) of physics education research conference proceedings using natural language processing. Of the ten themes found, none of them involved CT or CT education \cite{Odden}.  There is decent literature involving general computational thinking research, but it drops off as we get more specific to STEM and Physics. This sparseness presents an excellent opportunity for researchers to explore the various facets of physics computational thinking education. 

A report from the AIP Statistical Research Center investigated the initial employment of physics bachelors and PhD's in 2019 and 2020. 
According to the report, 46\% of new physics bachelors are employed, and 59\% of these are employed to the private sector.  Of those 59\%, 35\% work in an engineering field, and 24\% in a computer science field. Over 50\% of these engineers and computer scientists indicated that they use programming on a daily basis \cite{AIP}. Computational practices are expected for some of our physics graduates, and so we should be equipping them with skills they might need along the path to their degree. 

Introductory physics is in many ways an ideal space to begin learning computational methods while learning physics. In introductory physics classes, standard textbook problems are a primary means of assessment of students \cite{Hsu}. These problems are typically limited to those that can be solved analytically. Students are given idealized problems that often have internal assumptions made for them. While the reason for introducing problems in this way is to help better understand physics with basic examples, it has been inadvertently stifling physics education \cite{Fortus}. Students have the means and the access to cheat the current educational system. For example, students can use Chegg to answer book problems immediately. This means that checking student solutions based on answers alone is now (and probably has been) an inaccurate measure of learning. Not only is it an inaccurate measure of learning, it is perpetuating inequality in the classroom. A student can essentially purchase a good grade. Chegg does not have computational methods solutions as of yet. Chat GPT's introduction may even complicate computational assignments. A recent study by Kortemeyer found that in having chat GPT take an introductory physics course, chat GPT outperformed the other real students on the programming assignments \cite{Kortemeyer}. More work will need to be done to learn how we can leverage services like Chat GPT to supplement student learning.

Computational methods might force students to make their own assumptions explicit, and foster creativity. Odenn and Caballero produced a paper of a pilot study on computational essays. One of their results was that students reported that the computational essays helped facilitate creative investigation \cite{Odden-2}.   Computational methods allow exploration of different physics principles, and can address this early on in their physics career so that they may continue to build creativity and truth of messiness in physics throughout their education.  

 Expecting computational methods and thinking to be added to every introductory physics course may be unfeasible right now, but in the very least it can be argued that it should be added to the introductory physics courses for physics majors. Classes for physics majors typically have a smaller number of students. This means that instructors have more opportunities to provide individual support to them all, and learning these CT ideas potentially prepares students for their future careers \cite{AIP}.

\section{Literature Review}

\subsection{Computational Thinking in STEM}
There is an ongoing effort to integrate CT into many aspects of STEM curricula and degree programs. Samar Swaid examines the many aspects of CT and how they might be integrated into different STEM courses. Swaid takes careful note of what affordances CT can bring to each individual STEM discipline. For example Calc I and II afford a great environment for abstraction, data, and retrieving, but not so much for algorithms, design and evaluation. Alternatively, Biology I and II afford all of those aspects besides abstraction and design \cite{Swaid}. While the integration of CT is into STEM is generally agreed upon as the correct course of action, it is important to note that CT is going to look different for each discipline. This paper serves as a way to further define what CT looks like in the physics context.

A paper by Harangus presented a question to be solved using algorithms to secondary and higher education students, and found a relation between reading comprehension and problem solving skill. They note that many of the students did not even attempt the problem if they viewed it as too difficult \cite{Harangus}.   This work is important because it points out the motivation of students when tackling CT intensive problems. Physics, programming, and STEM overall are commonly associated with being high intelligence fields. This conception can be especially detrimental towards the attitudes of students taking STEM courses. In order to help retain students in STEM, the misconception of it being only for the intelligent students needs to be addressed. It is very important for STEM fields integrating CT to consider the attitudes of their students. CT problems should be introduced in such a way that eases students in. There should be a clear scaffolding of these CT concepts and practices so that students don't feel that the problems they encounter are too hard to even attempt.  

In a literature review of CT in STEM, Wang found that few studies examined equity associated with these courses \cite{Wang}. Most of the studies that addressed equity, did so focusing on various student groups like gender, socioeconomic status, and geographic location. There was one paper that focused on instructional strategies to increase equity.   Wang also found that while studies assessed student outcomes with CT integrated into the course, there were limited studies of assessments of the integrated CT course itself. In this paper, we do not directly address equity of CT in introductory physics.  We recognized that CT can be intimidating to students at first, but did not investigate how it may differ for different student populations. Our focus of this paper is on the learning goals of a CT integrated introductory physics class. We chose not to focus on equity because we did not want to do too many things at once and divide our attention.  Equity in CT in physics is important and needs to be researched carefully and wholeheartledy.

Weintrop et al in 2016 defines Computational Thinking for Mathematics and Science Classrooms. They say that with more traction of making CT a core scientific practice, a concrete definition of CT is needed along with a theoretical grounding for the form it should take in classrooms. They provide a definition of CT with four main categories: data practices, modeling and simulation practices, computational problem solving practices, and systems thinking practices. There are 5-7 subcategories within each. They did this by reading CT literature, interviewing mathematicians and scientists, and reading good CT instruction materials. They emphasize that the definition is CT in the science and math context. They created three lessons (physics, bio, chem) to implement their taxonomy, and state that implementing it in high school reaches a wider audience \cite{Weintrop}. Weintrop does a great job defining the computational thinking practices of a science course and even provides examples of how they implement these practices in a physics class. In designing our assessment, we will use this framework to connect physics learning goals to a specific CT practices.

We see from these papers on CT in STEM that there are many aspects to integrating CT into a STEM course. Weintrop focused on the practices that students should learn in order to develop their CT skills. Wang reviewed and discussed equity of CT in STEM finding that there were very few studies in this intersection. Although this is currently an understudied area, the importance of both CT and equity will push research in this direction when CT in specific disciplines is better understood.  Harangus found the relation between reading comprehension and problem solving skill. They also point out that problems that were seen as too difficult by the students were not even attempted. This can be an issue in physics. Lastly, Swaid discussed how different STEM courses afford different CT aspects.  With a general foundation of knowledge of CT in STEM, we can narrow down our focus to how these aspects are present in the physics domain.  

\subsection{Computational Thinking in Physics}
Caballero et al in 2012 implemented and assessed computational modeling in introductory mechanics. They claim implementing computation in intro physics courses has many benefits: modeling processes make complex problems tractable, and computation can explore the applicability and utility of physical principles. They also affirm that students who compute are doing work that is more representative of their potential future work as scientists and that learning how to debug code is part of learning computational modeling. They questioned the challenges students face when learning and applying computational problem-solving techniques, and how they can mitigate those challenges through instruction.
They found no statistical difference between performance of analytic vs computational homework. They also found that students with some previous programming were no more successful than ones without. They introduced computation to a large enrollment calculus based mechanics course at the Georgia Institute of Technology. Students were taught the computational language of VPython. \cite{Caballero}. The first thing we discuss about this paper is the two benefits of tractable complex problems, and physical principal exploration. Introductory physics education like we have said before is often limited in what can be solved analytically.  It is not that it can't be done or is too difficult, but that there is not enough time in a semester to spend the time effectively exploring these phenomena analytically.  Computational modeling provides the tools to tackle complex problems in a feasible and time efficient way. Along these lines, it also provides a place to play and explore. Often times analytical problems are difficult to change aspects of in a quick manner. Computational modeling allows quick changes of parameters so that students can explore more nuances of physical phenomena.  Again, it is not that this cannot be done analytically, it is just that computational modeling provides a feasible and time efficient way to learn and explore physics.
   
Obsniuk et al in 2015 did a case study on novel group interactions through introductory computational physics. They questioned what computation of physics in a group setting looks like. Their research was on an extended version of M\&I (Matter and Interactions), where students work with computational physics in a group setting.  This is called Projects and practices in physics or \(P^3\). They found two distinct strategies suited to computational tasks. They focus on the social exchanges between group members and the interactions between the group and computer. The group and computer interactions vary from actively sifting through code to observing a 3D display.  Students were to debug fundamentally correct code with wrong physical results. Bug recognition and bug resolution are the two necessary limits on physics debugging. When students found a bug, they blamed the error on their understanding of Python and not their understanding of physics. Students parsed through every line of code, asked each other if the line was correct and moved on. The paper classifies two debugging types as more strategic and less strategic. Self consistency was when they checked line by line and confirmed with each other (more strategic). Play or productive messing about was the less strategic way. Play shows the benefit of immediate visual display as a check where analytic tasks don't have that \cite{Obsniuk}.  This paper examines the intersection of computational physics and group work interactions.  Researching how students are interacting with each other and the programming environment is important because teamwork and adjacently communication are skills used on a daily basis for physics bachelors employed in the private sector \cite{AIP}. 

Weller et al. 2021 introduce a CT framework. They describe 14 practices that emerge from an integrated computational physics course: decomposing, highlighting and foregrounding, translating physics into code, algorithm building, applying conditional logic, utilizing generalization, adding complexity to a model, choosing data representation form, intentionally generating data, analyzing data, manipulating data, debugging, demonstrating constructive dispositions towards computation, and working in groups on computational models \cite{Weller}. Many of these practices that emerge in a computational physics course are practices that were found in Weintrop's paper.  There are a few that are not directly matched. For example, applying conditional logic is not one found in Weintrop's CT practices.  This may be an example where conditional logic is better suited in the physics context than the general STEM context. 

CT practices are getting better defined for physics in general, but there is still a lot of research to be done in many of the specific physics courses. The CT practices of an introductory course will be different from a capstone course. It may be that the practices are the same, however the sophistication of the practice would be different, which is still an important distinction.  We will now discuss the methods of our project and how we went about determining learning goals for introductory computationally integrated physics course.

\section{Methods}

\subsection{Participant selection}
To identify learning objectives in a computationally integrated introductory physics course, we conducted interviews with participants having some relation to the field, either within academia or within industry, beyond a bachelor's degree. Including both academic and industrial professions in physics provided a broader perspective about the role of computation.  Our subjects needed to meet at least one of the following criteria:
\begin{itemize}
  \item Is an active or past researcher on computational education in a physics classroom.
  \item Is an active or past instructor of a computational physics course.
  \item Is an active or past instructor of an introductory physics course.
  \item Is a physics graduate that works or worked in industry.
\end{itemize}


Using these search criteria, we identified and interviewed a total of 26 participants. Eight of these participants were physicists from industry while the remaining 18 were physicists in academia.  We did not collect any demographic information from our participants. While a diverse population is always preferred, we made a deliberate choice not to focus on the reporting of our participant demographic pool.  This may have provided an additional insight into how CT is defined for different populations within physics, however our goal is to determine the learning goals so that we may develop an assessment.  

\subsection{Interviews}

The interviews were semi-structured, about an hour in duration, and recorded and conducted online via Zoom. The interview questions focused on computation in the introductory physics classroom. For example, we asked, "What is the most important skill students can learn from a computational intro mechanics class?", and "What evidence would you look for to see if students met the learning goals in a computationally integrated intro mechanics class?" The full list of interview questions is provided in the Appendix. Of the 35 total interview questions, there were 24 that were asked of every participant, while there were a few that we only asked the industry or academic participants. For example, we asked industry physicists if programming is a skill they are expected to know, and we asked academic physicists about the classes they have taught before and how many times they have taught introductory mechanics. In total, industry physicists were asked 31 questions and academic physicists were asked 28 questions. Most questions we thought of ourselves, but there were three questions that stemmed from a theory of computational physics literacy described by Odden, Lockwood, and Caballero \cite{Odden-3}.  

Before starting the interview, we defined the difference between programming and coding (you must write code to write a program, but you need not write a program to write code), and then stated "When we say computation, it will typically be in the context of computation in physics." We chose to not explicitly define computation and computational thinking to the interviewees for a few reasons. First, computational thinking is hard to define and we did not want to bog down interview time with us trying to define it for them. Next, providing a concrete definition would force our narrative onto the interview participants and bias responses. Finally, a majority of the interviewees work closely or have worked closely with computation in physics, and so they responded based on their own interpretation of computational thinking. By leaving the definition open-ended, this allowed the interviewees to use their own interpretation and definition in their responses. Then, we were able to refine our own definition and understanding based on the interviews. 

\subsection{Analysis}

The interviews were first transcribed and then imported into the qualitative analysis software program Nvivo 12 Plus \cite{Nvivo}. Because our analysis methods involved \textit{coding} participant responses about program environment \textit{coding}, we distinguish the two interpretations of the word ``code" throughout the rest of the manuscript as follows: code/codes/coding\textsuperscript{q} will represent \underline{q}ualitative codes used in the analysis of participant responses while code/codes/coding\textsuperscript{c} will represent the \underline{c}omputational/programming side referenced by the interviewees. 


In this paper, we focus on our analysis of the five question topics that are most informative for the development of an assessment: \textbf{Important Computational Topics, Class Programming Environment, Limitations or Difficulties, Reasons for Computation in Introductory Physics, and How to Assess}. We drew on methods of constant comparison and grounded theory to generate codes\textsuperscript{q} that capture common themes related to these topics~\cite{Creswell, Glaser}. Glaser and Strauss describe four stages of the constant comparison methodology: comparing incidents, integrating categories, delimiting the theory, and writing the theory~\cite{Glaser}. They also describe that throughout the four stages of the constant comparative method, the researcher continually sorts through the data collection, analyzes and codes\textsuperscript{q} the information, and reinforces theory generation through the process of theoretical sampling. The benefit of using this method is that the research begins with raw data; through constant comparisons a substantive theory will emerge~\cite{Glaser}. Thus, this methodology not only shaped our interpretations of the data, but also influenced our data collection process. The research team went back and forth between collecting and analyzing qualitative data, comparing individual interviews to one another as they were being conducted in order to modify the interviews to focus on certain questions more closely tied to our research goals. For example, initially we asked questions about the programming environment "scratch", but these were found to be not as useful and were dropped from the later interviews.

Once all interviews were completed, the first author devised an initial coding\textsuperscript{q} scheme and coded\textsuperscript{q} all of the interviews. Then, the first and second authors met to discuss any disagreements in the way the interviews were coded\textsuperscript{q} and refine code\textsuperscript{q} definitions. These authors also discussed common themes that emerged from the interviews, and iteratively grouped the emergent codes under each theme. 

Once a final coding scheme was defined, we calculated inter-rater reliability with a third researcher by having the first author the third researcher go through fifty responses. Using a subset of twenty responses, we reached an agreement rate of 75\%. Then, the first author coded the rest of the data. 



\section{Results}

\begin{figure*}
  \includegraphics[width=6.5in]{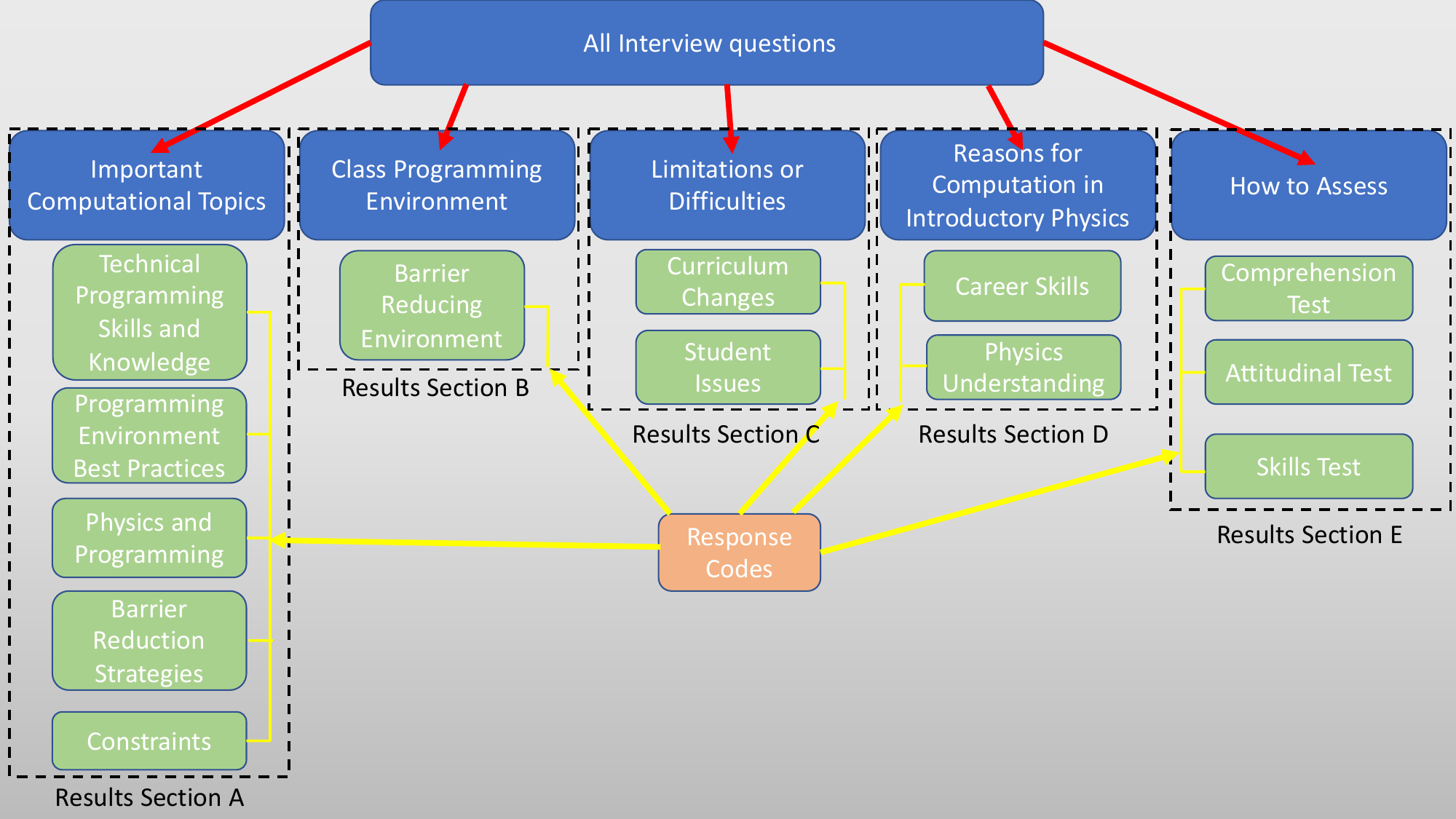}
  \caption{General outline of our results, in the order we present them. Of all the 35 interview questions, five are highlighted in this paper (represented in blue boxes). For each of the five questions, themes were extracted based on the participants' responses (represented in green boxes). The orange box represents individual response codes\textsuperscript{q} that fit into each theme. Responses could fall under multiple themes within a specific question.\label{fig1}}
\end{figure*}

This section is organized around each of the five questions of interest (see Fig.~\ref{fig1}), with an additional section about simulations at the end. We begin each sub-section with a description of the interview question itself and then describe the themes and codes\textsuperscript{q} identified from the interviews, with examples.



\subsection{Important Computational Topics}

This question elicits computation-specific topics or ideas in introductory physics that either the research team or the respondents themselves deem important. We identified 44 different response codes\textsuperscript{q} for this question. These response codes\textsuperscript{q} fell under five broader themes: \textit{Technical Programming Skills and Knowledge}, \textit{Programming Environment Best Practices}, \textit{Physics and Programming}, \textit{Barrier Reduction Strategies}, and \textit{Constraints}. 
Response codes\textsuperscript{q} could fit within multiple themes, but we place the response code\textsuperscript{q} with the theme we believe it most aligns with even though this categorization is not necessarily exclusive.

\subsubsection{Technical Programming Skills and Knowledge}
This theme encompasses technical knowledge and skills believed to be used more often in the programming and computational science context than the physics context. The following individual codes related to this theme:
\begin{enumerate}
    \item \textbf{Read Code\textsuperscript{c}} This response code\textsuperscript{q} had the highest response rate out of all the response codes\textsuperscript{q} in this study. Twenty of the twenty-six participants (77\%) mentioned that students should know how to read code\textsuperscript{c} by the end of a computationally integrated introductory physics class. There was one participant that explicitly stated that students should not be expected to know how to read code\textsuperscript{c} by the end of the term. This participant wanted students to focus on the underlying numerical calculations performed. They wanted students to use the medium most comfortable to them, and so that meant that not all students should be expected to know how to read code\textsuperscript{c}.  Almost all participants that mentioned that students should be able to read code\textsuperscript{c}, also added that students should only be able to read code\textsuperscript{c} that they have seen before.
    Note that all participants were given pseudonyms to protect their identity. Ore provides a typical response from the interviews.
    
    "Ore: I would expect them to be able to look at a code and be able to read out the different parts. So you should be able to identify where in the code are objects defined, where in the code is your initial conditions that run the code, where you are doing your calculations, What kind of calculations this list code is doing, and then also be able to make some predictions about what they think the code should produce."

In Indiana's response, we can see how they talk about how it is unfair to ask students about "for loops" if they have spent the entire term learning and practicing with "while loops".

    "Indiana: Yeah, they should be able to read the code that was introduced to them in the course. So for example, I brought up the idea that some people won't introduce for loops. So if you hand them a for loop after they spent the whole semester only learning about while loops. That's not fair." 

This is an interesting point in the development of the assessment. We don't want to assess students with programming code\textsuperscript{c} that they have not seen before, but every class is going to be taught a little differently. There will need to be adjustments to the assessment with time to calibrate a base set of programming fundamentals that are universal to every class.

    \item \textbf{Data Visualization} Nine participants (35\%) mentioned that plotting is an important skill to learn in an introductory computationally integrated physics course. Data visualization, and data communication are good skills for all scientists as well as many other professions to know and learn. Data visualization also provides another representation for physical phenomena which can be helpful for student learning. Kerry points out that making and describing plots is an important skill that is not necessarily constrained to computational methods, while Lee describes how making plots and plot interpretation deserve a lot of attention in the course.
    
    "Lee: But yeah, to meet the minimum (learning goals) is lots of practice doing visual representation using the computer. So a lot of plots and a lot of interpretation." 
   
    "Kerry: One could argue to what extent this is specific to computation, but once you have a program that's producing results: actually making plots and understanding plots and being able to describe plots. That's a very important skill, although that's a skill that also applies equally well to experimental measurements."
    
     While data visualization and interpretation has been a part of physics, it seems that giving students more practice with the creation and interpretation of plots within a computational environment is a path to be explored more. 
     
    \item \textbf{For Loops and While Loops} Eight participants (31\%) explicitly mentioned for loops or while loops as important skills to learn in an introductory computationally integrated physics course. Looping statements are typically the backbone in visualization of physical phenomena in motion or throughout time.  A similar idea of \textbf{Iteration}  was mentioned by six participants (23\%) as important, however we kept these ideas as separate codes\textsuperscript{q} because for loops and while loops are used specifically in programming environments, while iteration is a general concept used in CT. Mason talks about how it is important to know which parts of a program need to be inside of a loop or not.
    
    "Mason: Since almost all these calculations involved iteration, can they understand how a loop works. Can they make us a small prediction about the functioning of the loop, can they explain why or why not you could move some lines of code out of the loop or not."
    
    Something that was not nearly mentioned as much was the idea of computing time. There are plenty of variables that can included in a looping statement that don't change. It does not make the program wrong or produce an error, but it does increase the computational time required to finish the program.  We postulate that most introductory physics are not introducing programs that are computationally intensive enough to affect the computing time in a noticeable way.    
    \item \textbf{Writing Code\textsuperscript{c}} Seven participants (27\%) mentioned that writing code\textsuperscript{c} is important, but did not always specify conditions (i.e writing from scratch or writing from minimally working programs).  Kerry describes writing programming as being important by relating it to playing the piano.
    
    "Kerry: I think with anything, you need to do some writing in order to really understand it; to really understand what's happening there. One could make the same argument with playing a piano, right, that you might be able to read the music. That's a useful skill, but that doesn't mean you'd be able to play the piano. So if you want to play the piano, you need to actually play some notes right. We're not trying to make great pianists out of the intro mechanics students, but they should be able to do more than just read the music. They should be able to play a few notes."
    
     Kerry essentially says that students should be able to write some programming code\textsuperscript{c}, and that it is not enough to just be able to read it.  They mention that reading is an important skill, but it just is not quite enough for these introductory students.
    
\end{enumerate}  

  These four response codes\textsuperscript{q} had the highest response rate \textbf{Read Code\textsuperscript{c}} (77\%), \textbf{Data Visualization} (35\%), \textbf{For Loops and While Loops} (31\%), and \textbf{Writing Code\textsuperscript{c}} (27\%). There are a few other response codes\textsuperscript{q} that we found interesting, but not interesting enough to have their own sub-section. \textbf{Euler method} Six participants (23\%) mentioned Euler or Euler-Cromer methods as a skill for students to learn. The Euler method is an approximation method typically used in introductory computational physics because it is easy to use and provides a good enough result when modeling physical phenomena. Four participants mentioned \textbf{Execution Order} and \textbf{Functions}. \textbf{Execution Order} is comprised of responses involving how computers interpret commands in a certain order (first this, then that, etc). \textbf{Functions} was interesting because four participants advocated for the inclusion of functions as a skill to learn, however it is almost entirely counteracted by three participants advocating that students should not learn functions (in constraints section). When designing the assessment, we will need to be sure to limit the number of functions used.  This sort of makes sense because while some courses may use the same programming environment, they may not necessarily use the same functions. Other response codes\textsuperscript{q} that are within this theme but did not have as many responses were: \textbf{Conditional Statements and Arrays}, \textbf{Debugging}, \textbf{Spreadsheet Manipulation}, \textbf{Numerical Integration}, and \textbf{Step Size}.
\subsubsection{Programming Environment Best Practices}
This theme encompasses the best practices used in programming environments. These involve cleanliness, communication, and structure within and of programs.
\begin{enumerate}
  \item \textbf{Code\textsuperscript{c} Commenting} Nine participants (35\%) mentioned that students should comment code\textsuperscript{c} and should know how to comment code\textsuperscript{c}. This is a multifaceted and deeply rooted goal because it covers reading and understanding the code\textsuperscript{c}, and effectively communicating the physics within a program. There was one participant who stated that students should not comment their code\textsuperscript{c}, however this participant then followed up to say that the code\textsuperscript{c} should explain itself with meaningful variable names and not need comments. Aiden discusses how program documentation is important and that communication is a big part of this process as well. 
  
  "Aiden: And I think both of those are important. I mean, I think that like documenting the code is important for any kind of programming. And then I would hope that this would be ideally like built into the structure of the class where it would involve communication and collaboration as part of that."
  
  \item \textbf{Meaningful Variable Names} Eight participants (31\%) mentioned that students should either know how to create and assign variables or that students should make meaningful variable names. Hunter describes well written code\textsuperscript{c} as being able to speak for itself. They talk about how commenting is still a helpful practice, but in the end the code\textsuperscript{c} and variable naming should be all that is needed to understand what is going on.
  
  "Hunter: But definitely if you're working part of a large team if you're developing software that's going to be used by, you know, hundreds of people in perpetuity for many years, then I think it's critical. The last thing I would say is that I think commenting is not always the mechanism for making sure that code is understandable, and well written code can speak for itself, especially in a language like Python, which is a very nice naturalistic syntax. Surely commenting often does help and enhance the codes ability to stand on its own."
  
   \item \textbf{Intelligible Code\textsuperscript{c}} Two participants specifically from industry mentioned that student code\textsuperscript{c} should be intelligible: it should be clean, neat, and easy to follow. One participant stated: 
   
   "...it's basically that someone could come in after you and understand your code... I mean, I have PhD's where we still can't figure out what they did in their code right, and they don't seem to care. I mean, because they understand it, they don't care that their teammates can use it. So if you learn from the very beginning to have proper code hygiene, you know, comment, everything should be logical. Someone else should be able to follow this because it's in a logical sequence. I think it becomes natural for you and you'll always be you're always write code that at least someone will has a chance of following without having you sitting there next to them." 

   This participant notes that proper code\textsuperscript{c} hygiene involves a logical sequence.  This participant from industry describes how sometimes PhD students do not follow code\textsuperscript{c} hygiene and how it can be detrimental to the team if the program cannot be passed on to someone else without the program's translator. 

\end{enumerate}
\subsubsection{Physics and Programming}
This theme heavily focuses on physics and how physics intermingles with computational programming environments.
\begin{enumerate}
    \item \textbf{Identifying Core Physics} Nine participants (35\%) found it important for students to be able to identify the core physics statements within a computational program. We show two participant excerpts from Zion and Lee respectively:

    "They should be able identify where it (the physics) is. And but so again it depends how you designed your course. If it is designed with 'I want to teach them physics and this is a tool,' then they should know how to use the tool... I know how to use a car, but I don't know how every bit of a car works and there is a gas pedal and a steering wheel right that I know those really well..." 

    Here Zion describes the use of programming environments as a tool for understanding physics. Zion relates this relationship of tool understanding to the operation and use of cars. Zion says that they don't understand all the inner workings of a car, but essentially that they don't need to because those inner workings are unimportant. The real importance is in getting to the location Zion wants to be.  This analogy in this context, we understood Zion as meaning that as long as students know and understand the physics in the program, the rest does not matter as much.  

    "Providing minimally working codes or maybe minimally not quite working codes. It eliminates all of that barrier to just even getting started and then it's just like, Okay, concentrate on these lines right here, the physics is all in here. Do the physics. Right. And yes, you do have to encode it, but you know, take a look at the equations. We've written down. And, you know, there's the translation that translation right there is not easy I think for students and so for that to happen. Concentrate on something that's not easy. I mean, that's what you want them to do. You don't want them to say, Oh, I got a comment this and. Oh, I missed a colon or something like that. No. Put the cognitive burden, where it belongs."

    Here we see Lee talking about minimally working programs. Lee goes on to say that students cognitive burden should not be on the technical underpinnings of program syntax, but instead on the physics.  We believe that Lee is advocating the importance of focusing on the physics within a program out of everything else to make it easier for students.

    \item \textbf{Modify Physical Systems} Eight participants (31\%) want students to at the very least be able to modify physical systems computationally after taking this course.  This response code\textsuperscript{q} ranges from things like changing numeric values in a program as Kerry states, to extending existing programs as Mason states.

    "Kerry: I think it is important and and useful. And I think there's not much lost in a fair amount and gained by having them see code and interact with code and change lines of code and then run a program. Um, but they're not going to become good programmers. That would be an unrealistic goal to say they're going to become good programmers as a result... So that's where they're taking some code that exists and they're running it and they're they're analyzing the results and they're understanding something from it and they're changing parameters. Yeah, that's very appropriate."

    "Mason: ...And really understand enough of the model that they can extend it. So, for example, asking them to calculate and plot kinetic and potential energy means they have to understand What the constructs in the code mean, how they would calculate kinetic energy or potential energy of the system, and then how can they plot it. Now for that part I let them look at their old programs so they can pull out graphing code or energy calculated code or whatever they want."

    \item \textbf{Physics-Program Translation} Seven participants (27\%) advocated that students should be able to translate program notation physics into some other form/representation physics. Aiden describes how program notation physics is just another physics representation to learn. They talk about how it adds to their other representations like equations, graphs, and physical phenomena description. This is a place where reading comprehension may come into play. Is the reading and analyzing of a program considered reading comprehension or is it something else? This is similar to reading comprehension of a graph. Are graphs read or are their data interpreted? For now we will consider the analyzing and interpretation of programs as reading comprehension.
    
    "Aiden: In the same way that in any intro mechanics class like we would want students to be able to translate between you know description of a situation and equations and graphs and like translating among these different representations. I think that computation would add one more kind of representation. So I think a goal would be for students to be able to take a physical situation and figure out how to translate that into like an algorithm or into a program that they could run and vice versa, to be able to to look at a program and figure out how that translates into to the physical situation." 
    \item \textbf{Justification} Five participants (19\%) advocated that students should be able to justify their programs output. This concept is not specific to CT as justification is an ever present skill to be learned in physics and STEM. In Riley's quote we can see how it is more important to them that the students can explain and justify the results and output of the program. This notion is also seen in experimental physics labs.
    
    "Riley: I want them to explain you know what what the program does. I don't really care about the colon and all that stuff. Just tell me what the program does, and tell me how you know it's legitimate. That's why when they turn it in, I have them make a video where they show me their program and they just, you know, show me how it works and show the output and show that's legitimate." 
     
\end{enumerate} 

Some other response codes\textsuperscript{q} in this theme were: \textbf{Output Prediction}, \textbf{Authentic Programming Practice}, and \textbf{System Assumptions}. Respondents mostly focused on students being able to read code\textsuperscript{c} and find the physics within the program. Along with finding the physics, participants wanted students to translate between program notation physics and some other representation. They also expect students to be able to modify a program in some manner and be able to justify or explain the output of programs.  

\subsubsection{Barrier Reduction Strategies}
This theme represents practices or skills added to the course that are more focused on making CT easier for students to begin working with.
\begin{enumerate}
    \item \textbf{Scaffolding Incomplete Code\textsuperscript{c}}  Twelve participants (46\%) mentioned including code\textsuperscript{c} writing as a skill for students but only prefaced by the fact that scaffolding be in place. 
    
    "Gale: I firmly believe that the way you learn to write code is first writing, you know, filling in incomplete code. You know, filling in the blanks for incomplete code is scaffolding to being able to write the whole function or sentence or whatever yourself. It is much less intimidating than starting from a blank screen."

    "Lee: Providing minimally working codes or maybe minimally not quite working codes eliminates all of that barrier to just even getting started. Then it's just like, Okay, concentrate on these lines right here, the physics is all in here. Do the physics. Right. And yes, you do have to encode it, but you know, take a look at the equations we've written down. And, you know, the translation right there is not easy for students. I mean, that's what you want them to do. You don't want them to say, Oh, I got a comment this and oh, I missed a colon or something like that. No. Put the cognitive burden, where it belongs."

    With scaffolding, we can see that participants really care about students not being overwhelmed. It serves at least two purposes: 1. to make programming less scary for students, and 2. to focus student's attention on the part of the program that is important which is the physics.  What we take from this is that any program writing a student is doing in this course should be set up in such a way that students are only adding or editing the physics concepts. This way the focus is on learning physics and not program syntax.
    \item \textbf{Learn Real Applications}
    Six participants (23\%) advocated for students to learn about, but not actually practice, real computational physics applications. The reason why they don't want students to practice these applications, is because they are too complicated for an introductory class. Dakota talks about the value of learning some real applications. 
    
    "Dakota: I think if you're going to put that in computational physics or computational component to the introductory physics courses, it might be worth spending a lecture to talking about the range of types of computations, even if they're not going to use the full range, you know, so there are analytical computations, but there are there are other things. There are Monte carlo's. Okay, so they are not that they're going to do a Monte Carlo calculation or programming up but the only know what a Monte Carlo calculation is and when you apply it. nowadays neural nets are incredibly popular especially in the industry. So they ought to know what a neural net is and and how it works and why you would apply it, so there there are a number of computational tools in the toolkit." 
\end{enumerate}

    Some other codes\textsuperscript{q} that fit this theme were \textbf{Pseudocode\textsuperscript{c}}, \textbf{Experience Error Messages}, and \textbf{Exposure Comfort}. \textbf{Pseudocode\textsuperscript{c}} is the idea of writing a program outline outside of the program environment.  Typically this is done with pencil and paper. When participants mentioned pseudocode\textsuperscript{c}, they described it as a way that presented CT and programming as a lower risk, lower fear way to get students thinking computationally. \textbf{Experience Error Messages} had a couple responses where participants wanted students to get used to seeing error messages that arise from programming. The hope was that students could build emotional resilience to the common frustrations involved with errors encountered while programming.  \textbf{Exposure Comfort} had responses about students getting comfortable with programming environments by just being exposed to it.  They talked about how using a programming environment as a calculator was a low risk way of getting them used to the way the environment and tool works.
\subsubsection{Constraints}
Many of the constraint theme response codes\textsuperscript{q} have already been mentioned. This theme focuses on ideas or skills that participants do not expect students to know or are limited in some way. Constraint often implies a negative tone, but know that is not our intention in this case.
\begin{enumerate}
    \item \textbf{No Code\textsuperscript{c} From Scratch} Nine participants (35\%) advocated that students should not be coding\textsuperscript{c} from scratch in an computationally integrated introductory physics course. Participants that mentioned this also often mentioned scaffolding as important.

    "Vesper: It makes a huge difference for student frustration level. So, especially at the intro level. I always give them some kind of structure where they they have a couple lines, they have to fill in, but they're given that template and there are students in the class who would be capable of coding from scratch, especially if you kind of build up to it, but That requires sort of a deeper understanding of code logic and having a deeper level of mastery of coding and I don't have enough time in class to give all my students those things. and some of them are chem majors who are not taking computer science. So Right. Um, I don't even attempt to have them write code from scratch in intro physics ever"

    Vesper says that they never have their students write from scratch ever because it involves a deeper level of understanding that they don't have time to delve into. They also mention that it helps with keeping the students from becoming frustrated.

    "Xoan: so we we give them Templates. It's not quite the right word. Incomplete programs to start from scaffolding. So some of the basic stuff is there already. But we insist on having them do the key physics statements that are in those programs, which in the case of a Python program is a significant fraction of the number of lines unlike many programming environments. but it's still they cannot write from scratch. Okay, and we were sort of given up that as a goal"

    Xoan mentions that they have given up on students writing code\textsuperscript{c} from scratch as a goal insinuating that at some point they tried incorporating that as a goal and later decided against it.

    Earlier in our important computational topics sections we showed that some participants mentioned writing code\textsuperscript{c} as important, but they did not mention whether it should be from scratch or not.  Assuming all the participants earlier were talking about writing from scratch (seven participants), these nine participants saying students should not write from scratch entirely counter them.

    \item \textbf{Programming is Supplemental} Three participants mentioned that programming is not the focus of the course and that it is purely supplemental to learning the physics. Mason bolsters the idea that physics is the main topic to learn and that computational methods are just a means to that end.
    
    "Mason: Any computation things they're doing should be in the service of deeper learning and understanding of physics." 

    \item \textbf{No Functions}
    Three participants said that functions should not be learned in intro physics. This essentially counters the four participants that said that functions should be included. 
\end{enumerate}

\subsection{Class Programming Environment}
This question had 109 responses, with all twenty-six participants contributing to it. This question represents the programming environment(s) or language(s) that participants believe should be used in the introductory physics course. This question is not broken up by theme because all responses fall under the same theme: the environment that has the lowest barrier for students. A programming environment is the application that the programming is conducted within, while the programming language is the specific way of writing instructions for the computer to run.

\subsubsection{Python and Vpython}
\textbf{Python} (20 participants 77\%) and \textbf{Vpython} (14 participants 54\%) had the highest response rate in this category. Python and Vpython are free, open source programming languages. Vpython is short for Visual Python, and has a focus on animation and visualisation of code\textsuperscript{c}, where as Python does not. While these two had the highest response rate, they are languages and not environments. Some environments that were mentioned that use these languages, but did not have as high of a response rate were:  \textbf{Glowscript} (6 participants), \textbf{Jupyter Notebooks} (2 participants), \textbf{Trinket} (4 participants), and \textbf{Spyder} (1 participant).  Glowscript and trinket are both web-based with no local file storage and use Vpython. Three participants describe why they advocate for Python/VPython: 

"I exclusively use Python and VPython. Just because it has become the most accessible language and most popular language. I guess all over the world. And it's easy to get help. It's easy to get set up with Python more than with other languages. So I guess I'd vote for Python and the Python." 

 "And, you know, most physics labs are equipped with PCs or Macs are, you know, tablets or laptops. So even if students don't have devices that they can do the computation on they can all these are codes that run in a few seconds. So, you know, there's no reason why they should not be able to do. And if you're using a browser based platform like Glowscript or you know Trinket there is no added problem or challenge of installing software. So, you know, so the hurdle is less or minimal"

 "Um, so for intro mechanics. Um, but the most likely one that I would expect to see would be using Glowscript, VPython on Because from a logistical point of view, it's easy, that it's just you running into web browser without having to download things and you get visualization built right in."

We see participants describing how these languages and platforms are more accessible to students as they do not need to deal with downloading software. This means that they can run computational programs on any device as long as they have internet access. This is a great option for students who might share a computer or that don't have a computer they can reliably use. There were two responses that advocated for not using Vpython or Jupyter notebooks. The main argument for them was that Vpython won't be used later on in their career and that Jupyter notebooks were too much to learn for introductory physics students.

\subsubsection{Spreadsheets}

\textbf{Spreadsheets} were the next highest response rate (11 participants 42\%). The most common spreadsheet environments mentioned were Microsoft Excel and Google Sheets. Participants described spreadsheets as serving as a non-invasive way to teach CT, and that most students have already used or heard of spreadsheets so there is less apprehension towards learning/doing physics in them. Another reason participants advocated for spreadsheets was that students could actively see iterations and how data changes with each iteration. They also note that students will likely be working with spreadsheets in their future so more experience does not hurt.  Participants Frankie and Paris respectively talk about how they use spreadsheets for CT. 

"...I do sometimes use Excel or Google Sheets, just because when you code and you use loops in Python or any other language, what happens in the loop is pretty much a black box. So you don't see the the intermediate numbers being generated step by step. So it's it's useful to demonstrate that through a spreadsheet. So, you know, for the Euler method for every delta t increment, you can see how the velocity changes or how the acceleration changes or how the position changes and whatnot..."

"I'm also partial to Excel programming. Because a lot of students may not have come in with previous exposure to Python or other programming languages. But it's far, far more likely that they've come in with exposure to Excel or other spreadsheets, not necessarily programming in it. They may have used it just to like organize their books or keep track of like business finance it your family finances or whatever. But at least they're familiar with that software so that when you start, you know, introducing ideas of iteration into it. It's there's there's a little bit less of an overhead to that."

\subsubsection{Language Exposure Non-specific}

The third highest response rate was that the \textbf{Language exposure be non-specific} (10 participants 38\%). Essentially, participants stated that as long as students were getting practice with a programming environment, it did not matter which one they used. Dakota says that programming environments change over the years so focusing on learning a language is not for the language itself, but learning the ideas associated with using a programming environment: Computational thinking.

"Dakota: Yeah, that's a tough one, if they if there's one language that's been used for the course then the answer probably is yes, but the reality is, languages come and go... I think I tell people you know the language is going to change the language that they're going to use five years from now isn't going to be any of those. There's no reason to get too caught up in in the specifics of the language."

"Elliot: You know, but I mean, as long as you can program i don't i think picking up another language is not such a huge problem. Okay, right. I think it's maybe learning, you know, your first programming languages, you know, making sure that you have one"

 Some other similar response codes\textsuperscript{q} to this one were \textbf{Easy or Quick to Learn} (5 participants), and \textbf{Knowledge of Multiple Languages} (2 Participants). The \textbf{Easy or Quick to Learn} response code\textsuperscript{q} was again language non-specific with the stipulation that the language used be the easiest language to learn.  the \textbf{Knowledge of Multiple Languages} response code\textsuperscript{q} described the idea that students should learn more than one language in introductory physics. Their reasoning is that the more languages students learn, the more computational thinking strategies students can learn and can be applied to any language.

\subsubsection{Other Languages}

There were other languages that were brought up, however none of them were brought up as frequently as python or Vpython. The next highest language was \textbf{MATLAB} (5 participants) but every other language had three or fewer participants. There were however three participants that mentioned that students should not learn C/C++ and three that mentioned that students should not learn Java. Both of these environments were described as being overly complicated compared to Python/Vpython.  

\subsection{Limitations and Difficulties}
This question represents the struggles of adding computation into an introductory physics course. Knowing what to expect in terms of roadblocks to integrating computation into introductory physics is important so that aspiring instructors know what things to look out for. This question code\textsuperscript{q} had 21 participants contributing responses. These limitations and difficulties range from curriculum changes that need to be made to student attitudes towards computation.

\subsubsection{Curriculum Changes}
This theme has the two most cited difficulties: \textbf{No Room for Computation} (14 participants 54\%) and \textbf{Curricular Overhaul} (8 participants 31\%).  The most cited difficulty is that the introductory physics class does not have the space to learn computation as well as physics. Of those 14 participants, three of them were from industry so it shows that industry physicists are even thinking of this. How can computation be added to an already full content physics course? This difficulty, while a valid concern, was also addressed by most of the participants that mentioned it. They mentioned the idea that some things would have to be removed from the course in order to add computation, however the things removed are gained back in a better fashion through the implementation of computation. Jesse talks about this: "...Another limitation that gets brought up is that integrating computation will come at the expense of other material, sort of, you know, time on task spent solving analytical problems or covering additional concepts or what have you. I also kind of feel like that's bunk because if you integrate computation in the right way, you don't get so much of that loss and in fact it can it can simplify certain aspects of the physics teaching and learning like planetary motion. For example, it's a lot easier to simulate planetary motion and explore Kepler's laws through, you know, a computational simulation than it is to sort of go through the full complete analytical derivation and all of it's intense glory."

The second difficulty is the response code\textsuperscript{q} \textbf{Curricular Overhaul}. This is the idea that in order for computation to be effectively introduced into the introductory physics course, an entire overhaul of the curriculum is needed to integrate most of the content with computation. 

"Aiden: It's to do computation. Right. It's not something that can be done with just like, you know, throwing in this one assignment here. I mean, it's something that in the classes that do it well like they really build it up over the course of the whole semester. So it kind of feels to me like more of a sort of all or nothing that if I'm going to do it, I need to really like overhaul the whole class to do it, which is not something that I've done yet. I mean, if there are ways to do it in smaller doses then I would do that."

Just like how adding simulations to the course does not make the course a computational course (see section F of results), adding some computation without overhauling the entire curriculum is not an effective way of making the course computational. This then requires that in order to transition from an introductory physics to a computational introductory physics, instructors will need to spend a good portion of time overhauling their curriculum to include computation in most aspects of the course. This is a very time consuming task that is not always supported financially.

\subsubsection{Student Issues}
This theme focuses on issues that are more specific to the students than the class or curriculum itself. The three highest response codes\textsuperscript{q} in this theme are: \textbf{Student Rejection} (7 participants 30\%), \textbf{Differing Levels of Computational Literacy for Incoming Students} (6 participants 23\%), and \textbf{Technology Accessibility} (4 participants). 

\textbf{Student Rejection} talks about students will not accept computation in the course. Here are some excerpts talking about this:

"There are people though, who I think there are students who would resist it, because they would think to themselves, well I sign up for a physics course, not a computational course. So there's you definitely have to be careful about the context and the support"

"But there's also a lot of maybe like fear or anxiety around programming. That like when students are given code if they haven't seen it before, then that can be I guess like a barrier for them."

"The only difficulty there, especially with the pre meds, is that the pre meds have to pass this thing called an MCAT exam. And they are expecting a physics problem solving bootcamp when they wander into our classes and anything innovative like using a spreadsheet or programming language to solve problem isn't going to help them because they're purely there to pass an MCAT exam."

In the first quote we see the participant talking about how students would resist the course because the students view computational practices and physics as separate and non-overlapping things. The second excerpt talks about how students have a fear or anxiety towards programming. This fear is described as a barrier for students, and that barrier can lead to students shutting down and rejecting CT practices. The third excerpt is interesting because they discuss the MCAT. The MCAT (Medical College Admission Test) is a standardized, multiple-choice examination that is a prerequisite to the study of medicine \cite{MCAT}. They discuss how these students who are only taking a physics class to prepare for the MCAT are not interested in computational methods. Does the integration of computational methods and CT into an introductory physics class have any effect on the physics portion of the MCAT for these students? Research will need to be done to shed light on this area.

Along with student rejection comes the idea of \textbf{Differing Levels of Computational Literacy} for incoming students. Every student has a different knowledge level when entering this course. Because computational practices are becoming more prevalent even at the high school level, some students have worked with computational environments before they reach introductory physics while others have not. Just like how students have differing levels of physics knowledge coming into the class, practices will need to be put into place to make sure that any student regardless of background and prior knowledge can excel in this course. Jesse provides their take on this: "Um, there's a limitation in terms of sort of preparation. I feel like this starts to touch on a bit of an equity angle as well. If you don't spend so much time teaching the students computation, then those who come in with computation already in their back pocket, who are typically the students coming from like these, these well resourced high schools or, you know, families with with parents who are engineers or programmers, that kind of thing, those students will have an inherent advantage. That I do think is a legitimate complaint and a legitimate limitation here. And something that the instructor of course themselves is going to have a hard time addressing Unless you know you offer students a whole bunch of extra sessions to brush up on their computational skills or the computational literacy, which most instructors don't have the time for. So really, that's something that has to kind of be addressed before the students even arrive at the university like what Norway is doing right now with their their integration of computation across the high school curriculum."

Jesse views this student limitation from an equity lens.  They also bring up what Norway is doing. Norway is integrating CT into their mathematics, music, social sciences, English and programming classes in an attempt to emphasize CT in a multitude of disciplines, and incorporating CT as a core subject itself \cite{Bocconi}. Caballero's 2012 paper found no statistical difference between performance of students who took the class with prior programming experience and students with no prior experience \cite{Caballero}.  It could be that this is true of other courses, and so it may not be an issue. Also if other countries follow suit with what Norway is doing, then that may fix this issue as well.  

The last response code\textsuperscript{q} in this theme is \textbf{Technology Accessibility}. This is not only a student issue, but a course issue as well.  In order to program, students will need a computational device or computer in order to do so. This means that either students need to have this device themselves, or the course needs to provide them. While computers are quite prevalent now, not every course/student has access to them so it important to provide them. This is probably not a common problem, but when it is a problem, it can be a major issue. 
"Uri: Well, one limitations of course is both hardware and software.
Students have to have hardware, they're going to work at home, they have to have hardware at home. Not all of them can afford it. Same with software."

Uri talks about how students will need to have the hardware and software at home if they are going to work on these computational problems. They talk about how we need to be cognizant of students financial situations.  Even if students can use a computer on campus, this may prove to be a burden for commuter students.

\subsection{Reasons for Computation in Intro Physics}
This question represents the reasons why computation should be added into the intro physics class. We don't spend too much time on this because it has already been discussed why computation should be added to physics classes in the introduction section. That being said, we will briefly discuss some reasons why our participants found it important.

\subsubsection{Career Skills}
16 participants (62 \%) mentioned the idea that having computation in intro physics is important for building \textbf{Career Skills}. Many of the participants believe that the skills learned in this class will be used in the students future careers.  There was also the idea of this being a \textbf{21st Century Scientist Skill} (6 participants 23\%). Of the 16 participants that contributed to this code\textsuperscript{q}, 6 of them were physicists from industry. As there were only 8 participants from industry, this becomes a little interesting. There is definitely bias because these physicists either work with computational practices or were computationally adjacent, however one even mentions that it is not necessarily a skill just for physics: 
"Yes, I think that would be very beneficial. Just to help even students making the connection between how a computer works which can be very useful for jobs is even outside of physics. It's if they go work in technology in a bank or they go work in other areas. It just helps to have that connection early on."

"Um, well, yes. I think it should be somehow because it's an important part of doing real physics. Computation ultimately is going to be something that even if you don't do it yourself, you're going to need to know. And know enough about it that you can appreciate what's going on and you know, maybe even have a chance to collaborate with people who that's what they do... You know, whatever branch of physics to go into is going to be very complex and likely going to need some kind of computation to really do."
This industry participant talks about how it is an important part of doing real physics. They go on to mention that even if students don't do computation themselves, they will likely be around it and so having the knowledge to be a part of those conversations is useful.  
\subsubsection{Physics Understanding}
The next few response codes\textsuperscript{q} mostly involve student understanding of physics as the reason to include CT. They are: \textbf{Understand Physics Better} (12 participants 46\%), \textbf{Modeling Physics} (11 participants 42\%), \textbf{Skills Useful for Later Physics Courses} (10 participants 38\%), \textbf{Allowance of Non-trivial Phenomena} (10 participants 38\%), and \textbf{Creativity} (7 participants 27\%).  Most of these involve bolstering physics concepts and practices within students.  It makes sense that the main reasons for adding CT is to further increase students physics understanding overall.

\subsection{How to Assess}
This question was asked near the end of the interview as a way to solicit ideas for the development of our assessment. The participant responses were broken up into three different themes: Comprehension Test, Skills Test, and Attitudinal Test.

\subsubsection{Comprehension Test}
14 participants (54\%) talked about how they would assess for learning of CT by having students either read code\textsuperscript{c}, comment code\textsuperscript{c}, or find the physics inside code\textsuperscript{c}.
"Frankie: That that is a part of the learning outcome that you know By what I meant by them being able to interpret what the code does is, and I assess it by looking at the comments that they wrote."

"Mason: So our minimum expectations are that students should be able to read and interpret a very short program that instantiates a physical model."

This seemed to be the most agreed upon method and is a measure of students comprehension more than anything else.  Another response code\textsuperscript{q} in this theme is predict and explain results (9 participants 35\%). This entails students predicting the outcome of a program or explaining the results of a program.

"Noel: I think it would be important to learn the assumptions one makes in setting up a computation. You know, it can make the difference whether the result is reasonable or doesn't make any sense at all. I think often it seems to be easy to come to the conclusion that because the computation gave you a result that that it must be true without trying to figure out ways to test them.  In a sort of like a design of experiments, you know, try different parameters make making you know adjustments and seeing if the expected change and results is consistent so that you can say, you know, something more about whether the results that you're getting is actually believable."

These responses track with what has been discussed so far. Many of the participants would assess via reading, commenting, finding physics, or justifying a program output.  Most of these have a focus of physics but in the context of a computational environment.  We have noticed an emphasis of the physics being important and the programming being supplemental to learning physics.  Assessing students in this way seems to meet all of their expectations of not having them too engrossed in the syntax of coding\textsuperscript{c}.

\subsubsection{Skills Test}
4 participants (15\%) said they would assess students by having them write and run a program. This assessment style is both comprehension and skill based. Here is an excerpt from Gale: 

"So I think it's you know it is a performance assessment. You give them a task and they have to encode that and send you the code and hopefully when you hit run it will compile and run. But yeah, you might be able to assess aspects of coding you know, in a written assessment, but No, there's no substitute for the real thing." 

Gale says there is no substitute, in terms of assessment, for the real thing which is writing and running a program. Many of these four participants also mentioned that the students don't need to write and run a program from scratch.  This skills test is harder for students holistically as they are combining many different aspects of CT and physics. On top of that, there are many different places where a student can go wrong and make it so that their program does not run. It is not great for partial credit in that sense because if the student gets stuck debugging part of their code\textsuperscript{c}, they can't exactly move on until it is fixed.  Parsing through student code\textsuperscript{c} to find their issues is also no easy task as well.  It can be quite difficult debugging/grading multiple student computational assignments.     

\subsubsection{Attitudinal Test}
4 participants (15\%) said they would assess students by checking to see if the students have a better outlook or feeling towards computational practices. 
"Paris: So I guess what I would look for is just more of a classroom climate aspect. Like, does it seem to me that students are taken to this or excited about it or maybe the negativity like our students actually rejecting it or they at least kind of presenting themselves saying like yeah I guess this is what, this is what science does. So we have to do that.
You know, obviously I prefer them to be excited about it, but you know, considering the spectrum of people who go through these courses, I think that's maybe one thing to look for is just You know, is there active rejection or not. That's the hard thing to really, you know, measure and assess right and especially when It's especially when most of classroom assessments are geared towards you know content assessments."

Paris talks about how they care most about how students feel about CT and physics together. They talk about how they hope for excitement from the students. This is important because excitement is a form of interest and that is related to persistence in physics \cite{Hazari}.  They talk about how most assessments are geared toward content and views attitudinal assessments as harder to administer because of this.  If we want physics to be an inclusive place, student attitudes become important to analyze. 

\subsection{Simulations}
One of the things we sought out was to determine whether simulations were considered computation or not. We found that while simulations are computational in nature (made from computational practices), they indeed were not considered one of the computational skills students should learn. One excerpt that we feel highlights what most respondents said was this one: 

"...especially with that simulation, you are very far away from what's happening under the hood. You don't see how the computer is actually working. For a [physics] major I think it's important to see that, you know, there's a few important lines of code here that implement basic physics that we want you to know about and you should be able to go in and make changes to a couple of lines of code in order to generate different results." 

Respondents often said that simulations are great for learning. It is an excellent learning tool that any student can use to supplement their understanding of physics \cite{Banda}.  The main point that was often brought up was that simulations were not enough for physics majors.  For physics majors, respondents wanted them to see, create, and manipulate the physics that would make a simulation. In most simulations, students don't have that kind of access. This was a very important point to our participants. There has been a push to include computation into physics classes, and so many curricula have added simulations to meet this criteria.  What we have learned from these interviews is that simulations are not enough if the course is for physics majors. Simulations are not bad, they are great learning tools and should be added to physics courses, but they do not give students the opportunity to learn the computational skills that most participants mentioned to be important from these interviews.

\section{Discussion}
The goal of the interviews was to discern what student skills or ways of thinking are important to learn after taking a computationally integrated introductory physics course as to come closer to answering (RQ1). We attempted to answer this by interviewing 26 physicists and asking them about CT in introductory physics. We used constant comparative method, grounded theory, and emergent coding\textsuperscript{q} on these interviews to pull out codes\textsuperscript{q} and themes. We analyzed and presented 5 question topics deemed important from these interviews: \textbf{Important Computational Topics}, \textbf{Class Programming Environment}, \textbf{Limitations or Difficulties}, \textbf{Reasons for Computation in Introductory Physics}, and \textbf{How to Assess}. Within \textbf{Important Computational Topics}, we found that reading code\textsuperscript{c}, data visualization, code\textsuperscript{c} commenting, identifying core physics, scaffolding incomplete code\textsuperscript{c}, and not coding\textsuperscript{c} from scratch were some of the most prevalent response codes\textsuperscript{q}.  Within \textbf{Class Programming Environment}, we saw that Python and VPython by far were the preferred languages to be used in introductory physics. We also saw that spreadsheets were also brought up frequently as an acceptable environment. Within \textbf{Limitations or Difficulties}, we saw that changing the curriculum was the most cited difficulty of integrating computational practices into the course. We also had a few responses that focused on the student perspective and how they might have difficulties with accessibility, differing levels of incoming literacy, and rejection of material. Within \textbf{Reasons for Computation in Introductory Physics} We saw that career skills were the most cited response with student physics understanding following after. Within \textbf{How to Assess}, we saw participants discuss ways on how they might assess CT in intro physics. They mentioned attitudinal tests, skills tests, and Comprehension tests. Comprehension tests had the most responses. 

In Weller's framework, we notice a few similarities like: translating physics into code, adding complexity to a model, analyzing data, and demonstrating constructive dispositions towards computation that were also found in our interviews \cite{Weller}. Debugging was something that only three of our participants mentioned. This has its own category in Weller's paper so we were surprised to see that it was not mentioned more in our interviews.  Other CT practices were not explicitly mentioned either. Highlighting and foregrounding was not mentioned by our participants explicitly, however our participants did describe some processes of highlighting and foregrounding like making system assumptions, understanding execution order, and practicing pseudo-code.  

One overarching theme we notice from the interviews is the idea of students reading and understanding program code\textsuperscript{c} as a learning goal. In the interviews, codes\textsuperscript{q}, and themes, we see that students reading code\textsuperscript{c} is the most agreed upon skill to learn. Reading code\textsuperscript{c} permeates through many aspects of the response codes\textsuperscript{q} from the interviews. Another overarching theme we notice is the general barrier reducing strategies mentioned. Many participants mentioned that CT in physics can be daunting. They brought up ways and strategies instructors could attempt to mediate this via scaffolding incomplete code\textsuperscript{c}, focusing on physics rather than computational environment syntax, and using computational environments and languages that are more accessible. The learning goals are still the same introductory physics learning goals. Additional learning goals for CT in introductory physics are needed though, and we have seen that they mostly take the form of computational reading comprehension. 

We would be remiss not to mention that part of the reason that many participants talked about students reading code is because we specifically asked them if students should be able to. It is important to note that it was explicitly brought up in the interviews by the interviewer. Most of the questions were open-ended and not necessarily prompted about a specific facet of CT. For example we asked participants about programming environments in general and not "python" or "spreadsheets". All of their responses (to this question) are more valid in the qualitative sense because their responses were not steered by the interviewer. Even though participants were explicitly asked about reading code, we believe the results are still valid because aspects of reading code were present in participants answers of other questions. We hypothesize that if we did not ask explicitly about reading code that it still would have come out of the interviews thematically. 

Twelve participants mentioned something about scaffolding incomplete code\textsuperscript{c}, nine participants mentioned not coding\textsuperscript{c} from scratch, eight participants mentioned that students should be able to modify physical systems, and seven participants mentioned that students should be able to write code\textsuperscript{c}. What we can see from these interviews is that participants generally expect that students should be able to use a programming environment to in the very least edit a program with physical phenomena. Almost all of these participants were very adamant about how students should be writing their code\textsuperscript{c}. Some reasons were that they were very conscious of how coding\textsuperscript{c} and programming were perceived by students.  They recognized that programming and coding\textsuperscript{c} are viewed by students as scary or only for computer people, so to reduce this sentiment, they focus on having students only work on specific parts of the program rather than the program as a whole.  This is often seen as better for students since they can essentially avoid the less fun and potentially deterring aspects of coding\textsuperscript{c} like syntax and debugging, and can in turn focus on the physics that they likely have more of a passion for. 

A lot of these responses from these four codes\textsuperscript{q} elude to minimally working programs. Minimally working programs in the physics context have been adapted slightly to fit their goals. Typically a minimally working program would involve a short program with one small focused bug to be found and corrected by the user.  In its adaptation in physics, it is now typically a full working program but with incorrect physics to be found and corrected by the user \cite{Pawlak}.

Another selling point that Mason makes is that coding\textsuperscript{c} in real practice does not involve coding\textsuperscript{c} from scratch. Often times we will be building off of someone else's work or our own and so starting from a blank program is rare. It makes sense that participants do not want students writing from scratch if it is a practice that they do not do themselves. Mason mentions that this is something that the pros do. 

To answer our two research questions, we see that the learning goals do not differ much between an introductory physics class and a computationally integrated introductory physics class. The main goal of the course is for students to learn and practice physical principles.  In the computationally integrated course, the way students practice is different since they are using a computational environment as a tool. While there are some base skills and practices that are more computational than physical (learning for loops and while loops, commenting code, creating variables), students are mostly expected to learn another representation of physics.  This suggests that our assessment should focus on the base CT skills needed for students to modify physics in a minimally working program, and focus on the intricacies of switching between various physics representations and the program physics representation.  From our interviews we see that this assessment should be written using the language of Python, Vpython, or spreadsheets.  The assessment should also be written and presented in a form that is less intimidating for students so they can continue to learn physics confidently.  

\section{Future Work}

Our future work involves a computational physics reading comprehension assessment. This assessment will be mostly multiple choice and multiple choice multiple response. In terms of determining how we can design an assessment for the computational learning goals found from these interviews, the most "concrete" learning goal we can test is reading code. This is also bolstered by most participants mentioning that they would attempt to assess CT in introductory physics via comprehension test. While skills tests and attitudinal tests are also feasible and need to be researched, based on this study we have determined that we will design a comprehension test via reading program code. 

The other forms of assessment that should be explored are attitudinal test, and the skills test. While this may not be in the scope of this project, it will be important to design an attitudinal assessment for student sentiments towards CT in introductory physics. The reason this is important is because computational science, CT, programming, and physics are associated with the nerd-genius stereotype in STEM as described by Christine Starr. Adding CT to physics makes sense for bettering student understanding and learning practical skills for their future careers, but it does not help us in the long run if it is also deterring and pushing out certain demographics. For example with the nerd-genius stereotype, STEM domains are often associated with people who are nerdy, like dungeons and dragons, are unattractive, are inherently smart, and like computers.  For women in particular, this can potentially push them away from the field. For example, historically, society as a whole has placed a higher value on women's beauty than men's. If society places higher value on beauty, and there is a nerd-genius stereotype with STEM which implies unattractiveness, then this likely contributes negatively to women's STEM identity which is related to motivation and persistence \cite{Starr}. It is important that we can analyze students attitudes on introductory computational physics so that we can make changes to create an inclusive and welcoming environment for any person who wants to learn physics.  
Another assessment that is not in the scope of this project is the skills test.  As we noted from before, there were quite a few participants that mentioned that writing, modifying, and running code (not from scratch), is an important skill for students to learn.  The skills test is important, because it is assessing students use of a programming environment tool to solve physics.  This form of assessment most closely matches what students are learning/doing in their computationally integrated introductory physics, however this form is the hardest to generalize across many classes nationally.  

 With these interviews complete, we will now work on a draft assessment of CT in introductory physics. The assessment, informed from this study, will be a computational reading comprehension assessment. We will focus on questions that show if students can read a physics program written in Vpython and can identify the physics principles involved, translate between algebraic notation and program notation, and can explain if results of a program are reasonable physically. Because we wish to focus on student comprehension more than a skills test, the assessment will not involve the creation, editing, or manipulation, or extension of a program in a programming environment.  Instead the assessment will take on a multiple choice style as to concentrate on the reading comprehension.   This assessment will need to undergo scrutiny and pass many validity tests. When the assessment has reached acceptable validity, we will then pilot the assessment to student populations.  
\clearpage


\clearpage

\hspace*{-2cm}\includegraphics{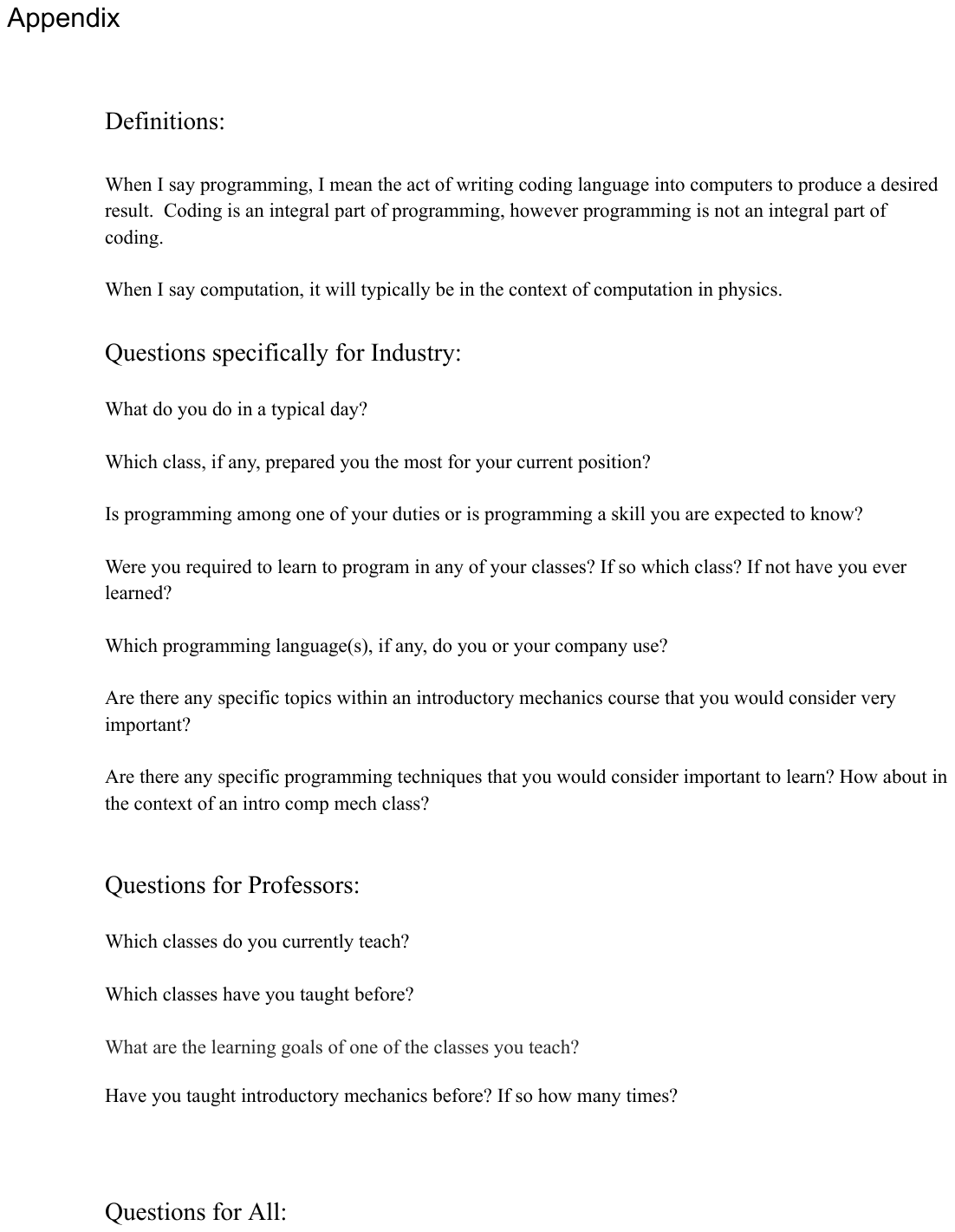}
\clearpage
\hspace*{-2cm}\includegraphics[page=2]{InterviewQuestions.pdf}
\clearpage
\hspace*{-2cm}\includegraphics[page=3]{InterviewQuestions.pdf}
\clearpage
\end{document}